\documentclass[twocolumn,showpacs,preprintnumbers,superscriptaddress,prl,floatfix]{revtex4-1}

\usepackage{amsmath,amssymb}
\usepackage{textcomp}
\usepackage{hyperref}
\usepackage{graphicx}

\renewcommand{\vec}[1]{\ensuremath{\mathbf{#1}}}

\begin{document}
\title{Lattice dynamics of anharmonic solids from first principles}
\author{O. Hellman}
\author{I. A. Abrikosov}
\author{S. I. Simak}


\affiliation{Department of Physics, Chemistry and Biology (IFM), Link\"oping University, SE-581 83, Link\"oping, Sweden.}

\begin{abstract}
An accurate and easily extendable method to deal with lattice dynamics of solids is offered. It is based on first-principles molecular dynamics simulations and provides a consistent way to extract the best possible harmonic -- or higher order -- potential energy surface at finite temperatures. It is designed to work even for strongly anharmonic systems where the traditional quasiharmonic approximation fails. The accuracy and convergence of the method are controlled in a straightforward way. Excellent agreement of the calculated phonon dispersion relations at finite temperature with experimental results for bcc Li and bcc Zr is demonstrated. 
\end{abstract}

\maketitle

Ability to predict phase equilibria and structural transformations in solids under pressure and temperature is of vital importance to both science and industry. The density functional theory (DFT)\cite{Kohn1965c} allows one to calculate thermodynamic properties of solids from first principles, i.e. without any adjustable parameters or information from experiment. Though it is a ground-state theory the crucial effect of thermal vibrations of atoms can in principle be included via the harmonic approximation\cite{Fultz2010}. The DFT-based calculation of corresponding phonon dispersion relations is nowadays routine\cite{Baroni2001b}. The widely used so-called quasiharmonic approximation\cite{Fultz2010} for the free energy employing zero temperature DFT calculations of volume-dependent phonon frequencies is well developed\cite{VandeWalle2002}. However, it assumes that the potential energy surface does not depend on temperature and is limited to systems where anharmonic effects can be neglected. In particular it is known to dramatically fail when applied to crystal structures which are dynamically unstable at 0 K, i.e. their phonon spectra contain imaginary frequencies, but which stabilized due to the anharmonic contributions to the free energy at elevated temperatures\cite{Dubrovinsky2007}. 

The failure of the harmonic theory mostly arises from the 0 K Taylor expansion of the potential energy surface up to just the quadratic term, which at temperatures close to the melting point becomes increasingly inaccurate. Consequently, theoretical treatment of anharmonic effects has attracted great interest\cite{Klein1972}. The self-consistent phonon approach was formulated by Born\cite{Born1951}, and developed by Born and Hooton\cite{Hooton,Born}. Choquard\cite{choquard1967} and Plakida\cite{Plakida1970} addressed the problem employing the perturbation theory based on the so-called double-time Green functions. However, such methods are difficult to combine with DFT-based first-principles calculations. The expansion to higher-order terms within the standard techniques is cumbersome and currently unfeasible, in particular due to issues of numerical precision.

On the other hand, recently Souvatzis et al.\cite{Souvatzis2008a} have introduced the so-called  self-consistent ab initio lattice dynamics approach, which is conceptually similar to the so-called renormalized harmonic approximation\cite{choquard1967}. The theory introduces anharmonic effects through the temperature dependence of phonon frequencies mimicking interaction of phonons. The price is the additional approximations, such as fixed amplitudes of the vibrations and a limited phase space of allowed excitations. In Ref. \cite{Souvatzis2008a}  experimental dispersion relations for the group IV metals have in general been reproduced, though substantial error has been reported for a determination of transition temperatures between different phases \cite{Souvatzis2009}.

Fourier analyzing the velocity autocorrelation function from molecular dynamics (MD) represents an alternative approach to determine the phonon dispersion relations at finite temperature. It works for classical MD simulations\cite{Dickey1969}, but the system sizes currently accessible to \emph{ab initio} MD (AIMD) makes this approach prone to finite size effects and can thus not be used with meaningful accuracy. The common practical way of dealing with anharmonic effects in DFT based simulations of phase transformations is combine AIMD with a thermodynamic integration using the harmonic energy as the starting point\cite{Vocadlo2002,Grabowski2009c}. In the standard harmonic approximation, however, the imaginary phonon frequencies mean that the free energy is not defined and therefore the starting point is missing. Therefore often an artificial model with a fitted classical potential is used instead\cite{Vocadlo2002}.

We propose a method that intrinsically solves all the issues mentioned above. It uses \emph{ab initio} molecular dynamics simulations and provides a consistent and computationally easy way to extract the best possible harmonic -- or higher order -- potential energy surface at finite temperatures to study the lattice dynamics and thermodynamic properties of the system under consideration. At present we show that already the second order terms are sufficient to accurately describe systems where the anharmonic effects are known to be prevalent and as examples we present calculations for the strongly anharmonic bcc Li and Zr. We stress, however, that if the second-order approach proves insufficient for some complex system it is easily extended to any order. Further, our method allows one to control the accuracy and convergence in a straightforward way, in contrast to most of the previous methods.

For clarity the equations are derived for a monoatomic system but it is easy to extend them to the general case. Usually the harmonic approximation is given as a Taylor expansion of the potential energy surface (at 0 K) in atomic displacements $u$, truncating after the second order terms:
\begin{equation}
\label{eq:taylorexp}
U=U_0+\frac{1}{2}\sum_{i,j,\mu,\nu} u_\mu(\vec{R}_i) \frac{\partial^2 U}{\partial u_\mu(\vec{R}_i) \partial u_\nu(\vec{R}_j)} u_\nu(\vec{R}_j)+\ldots
\end{equation}
where the second derivatives of the energy surface $U$ with respect to displacements $u$ are called the force constant matrices, $D^{ij}_{\mu\nu}$. Here $\vec{R}_i$ and $\vec{R}_j$ are ideal atomic positions and $\mu$ and $\nu$ correspond to the Cartesian coordinates. The established way to find the force constant matrices $D$ is through the linear response\cite{Baroni2001b} or the small displacement method\cite{Alfe2009}. In these methods the crystal is considered in its ground state and displacements of individual atoms, either via perturbation theory or a supercell approach, are introduced. In this work however, we want to obtain the best possible harmonic potential for a given temperature and volume, derived from a thermally excited crystal.

In an expansion in the second order the forces are given by
\begin{equation}\label{eq:harmforce}
{\vec{F}}_i=\sum_j \vec{D}_{ij} \vec{u}_j.
\end{equation}
We want to find the effective force constant matrix $\tilde{D}$ that best represent the real forces. To do this we run $N_t$ time steps of \emph{ab initio} molecular dynamics simulations for a supercell containing $N_a$ atoms, and at each time step $t$ we store forces and displacements. We then seek to minimise
\begin{equation}\label{eq:minimize_f}
\Delta \vec{F}= \sum_{t,i} \frac{1}{N_t} |\vec{F}^t_i - \tilde{\vec{F}}^t_i|
\end{equation}
where $\vec{F}_{i}^t$ is the force acting on atom $i$ at time step $t$ from the simulation, and $\tilde{\vec{F}}^t_i$ is the force given by Eq.~\ref{eq:harmforce} with displacements taken from time step $t$. If the number of time steps are larger than $3N_a$ we have an overdetermined system of equations for $\tilde{D}$. We can then obtain the linear least squares solution of $\tilde{D}$. Having converged with respect to the number of time steps we have defined a force constant matrix at temperature $T$ we can obtain the phonon dispersion relations and free energy. 

\begin{figure}[!htp]
\vspace{3mm}	
	\includegraphics[width=8cm]{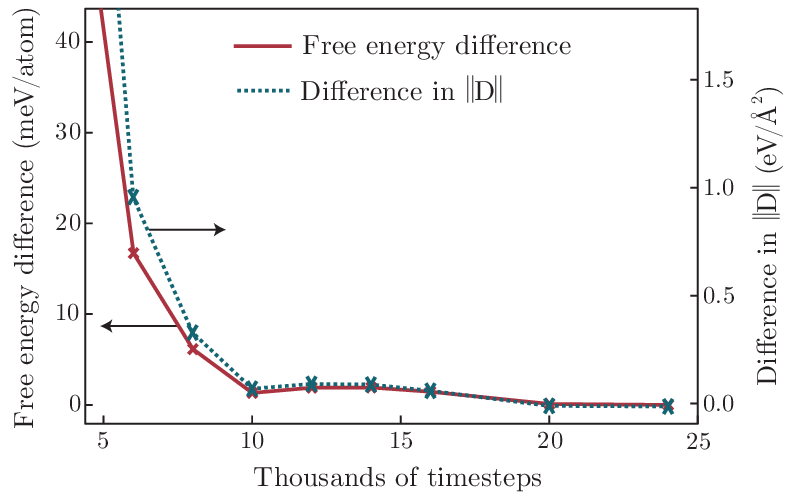}
	\caption{\label{fig:konvdata}(color online) Difference between the free energy (solid line, left axis) and $||D^{11}(\vec{R}_1^1)||$ (dashed line, right axis), the Frobenius norm of the nearest neighbour interaction derived from increasingly longer simulations for bcc Li at $T=300$K. As a reference the results obtained from simulations with 25000 time steps are used, which are considered to give the converged results (0.43 eV/\AA$^2$ for $||D^{11}(\vec{R}_1^1)||$).}
\end{figure}

To monitor the quality of the force constant matrix for bcc Li we determine it from successively longer simulations, and show the result in Fig. \ref{fig:konvdata}. It clearly demonstrates that the suggested scheme allows us to derive a unique and well converged effective force constant matrix $\tilde{D}$.

We explicitly note that the forces, displacements and force constant matrix in this example are obtained at temperature $T=300$K, whereas the quasiharmonic approximation would derive $D$ at 0 K and from that obtain the free energy as a function of temperature.
Note that in our method the force constant matrix $D^{ij}$ is defined per pair $ij$ in the supercell. However, a careful remapping allows one to define it for each lattice vector, that is
\begin{equation}\label{eq:fcmdef}
\tilde{D}^{ij}_{\mu\nu} \rightarrow \tilde{D}^{\alpha\beta}_{\mu\nu}(\vec{R}_i^\kappa)
\end{equation}
where $\alpha,\beta$ are indices to atoms in a suitable unit cell of choice (smaller or equal to the supercell), $\vec{R}_i^\kappa$ is vector $i$ in coordination shell $\kappa$ that contains only pairs of type $\alpha,\beta$. To increase numerical stability we use the point group of the coordination shells as well as symmetry properties of the force constant matrices. All the symmetrization procedures are similar to those detailed in Alf\`e et al.\cite{Alfe2009}, although used for a different purpose. The small displacement method uses as few displacements as possible and then constructs all force constant matrices through symmetry relations, putting high accuracy demands on the calculated forces. Our method yields all force constant matrices that are then forced to obey the symmetry relations of the lattice, reducing the need for numerical accuracy in the calculated forces.

In the spirit of Ref. \cite{Grabowski2009c} we analyzed the forces to asses the errors in our method. We analysed forces from a subset of configurations, calculated with different level of accuracy. $F^L$, forces with an accuracy typical for first principles MD simulations were compared to $F^H$ fully converged results%
\footnote{In this case $F_L$ are from a 128 atom bcc Li supercell, $2\times 2 \times 2$ kpoint grid, and $F^H$ from the same configuration, with 50\% higher energy cutoff and $7\times 7 \times 7$ kpoint grid.}
. We found that the norm ratio, $\langle|F^H|/|F^L|\rangle$, oscillated around unity ($\approx 1.0015$), with standard deviation of 0.007. This, as well as the results from several test calculations and data presented in Fig. \ref{fig:konvdata}, allows us to draw the conclusion that since the error in forces is normally distributed around the "true" value, our method of least square fitting of many configurations appearing during the MD simulations can reproduce force constant matrices with high accuracy.

The fact that we do not do a Taylor expansion around equilibrium at 0 K but produce the potential energy surface according to the least square fit of a quadratic form at a particular finite temperature allows one to apply our method to systems where the quasiharmonic approximation traditionally fails. We notice that the standard quasiharmonic approximation has a temperature range at low temperatures where it works well for dynamically stable systems. Our method moves this window to any temperature of interest. Although not explicitly anharmonic, it gives us truly the best harmonic fit to the fully anharmonic energy landscape and therefore contains the information about anharmonism implicitly through the temperature dependent vibrational frequencies.

At this point it should be stressed once again that this method allows for an easy, though a more time-consuming, extension to explicitly handle anharmonism by including more terms in the expansion in Eq. \eqref{eq:taylorexp}. 

We chose bcc Li and Zr to demonstrate the advantages of the proposed technique. Li, long thought of as a simple metal, has a complex phase diagram. It undergoes a phase transition from the bcc structure at room temperature to the close-packed R9 structure below 70 K\cite{Overhauser1984}. With pressure it behaves anomalously transforming to the low symmetry \emph{Cmca} structure\cite{Neaton1999a}. Quasiharmonic phonon dispersion relations for the bcc structure show that it is dynamically unstable at 0 K, and the free energy is not defined. Zr has long been used as a model system for martensitic phase transitions and is a well known example of a strongly anharmonic solid. The nature and origin of the stabilization of the bcc phase has been discussed in numerous previous studies\cite{Ye1987}.

All electronic structure calculations were carried out with the projector-augmented wave (PAW) method as implemented in the code VASP\cite{Kresse1999,Kresse1996,Kresse1993b,Kresse1996c}.
We used a 128 atom bcc supercell ($4 \times 4 \times 4$) for the MD simulations. For the BZ integration we used a $2 \times 2 \times 2$  k-point mesh and Fermi smearing corresponding to the simulation temperature. Exchange and correlation effects were treated using the Perdew--Burke--Ernzerhof\cite{Perdew1996} functional form. A plane wave cutoff of 140 eV was used for Li and 154 eV for Zr. Both systems were considered at their theoretical equilibrium lattice parameter.

We used a 2 fs time step, which is suitable for both systems, and set the temperature with the Nos\'e-Hoover thermostat\cite{Hoover1985,Nose1984}. To fully check the convergence of our method we ran the calculations for a total of 44000 time steps (88 ps) after the initial equilibration and extracted the force constant matrix at fixed time intervals. After about 40 ps the free energy was converged to below $0.5$ meV/atom (see Fig. \ref{fig:konvdata}), which is an accuracy exceeding that of the underlying DFT approximations.

\begin{figure}[!htp]
\vspace{3mm}	
	\includegraphics[width=8cm]{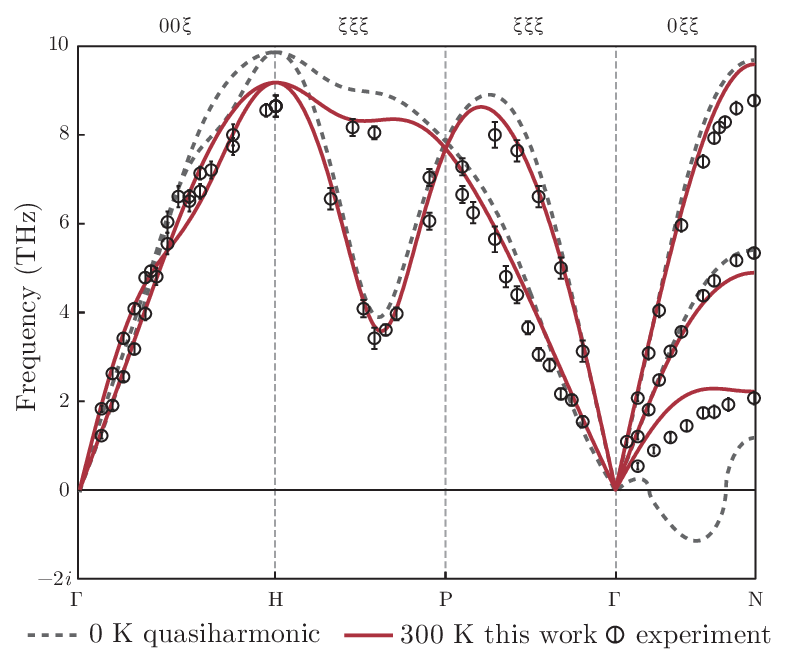}
	\caption{\label{fig:li_disprelations}(color online) Phonon dispersion relations in bcc Li along high-symmetry directions. The symbols are experimental values (293 K)\cite{Beg1976}, the solid lines correspond to calculations carried out at 300 K with the method proposed in this work and the dashed black lines to the 0 K harmonic calculations.}
\end{figure}

The success of the suggested method on the quantitative level can be judged from a direct comparison of the calculated phonon dispersion relations (easily extracted from the calculated force constant matrix) to experimental. In Fig \ref{fig:li_disprelations} calculated dispersion relations at 0K, room temperature and experimental results are shown for bcc Li.
The 0 K dispersion relations, obtained from the quasiharmonic approximation reveal imaginary frequencies along the $\Gamma-N$ direction. Using our method at finite temperature all imaginary frequencies disappear and we have an excellent agreement with experimental values.

\begin{figure}[!htp]
\vspace{3mm}
	\includegraphics[width=8cm]{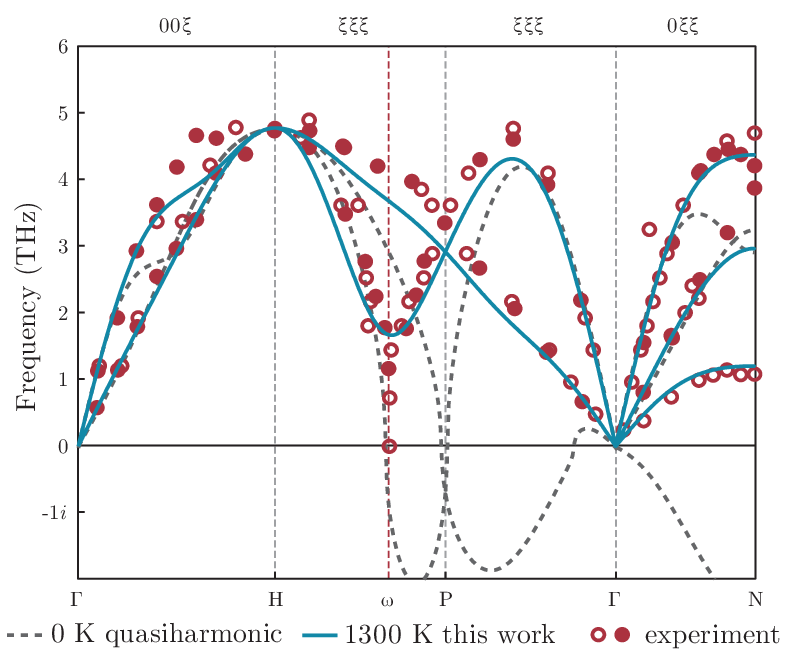}
	\caption{\label{fig:zr_disprelations}(color online) Phonon dispersion relations for bcc Zr. Solid lines correspond to calculations at 1300 K, dashed lines to the quasiharmonic results and symbols to experimental values from Heiming et al.\cite{Heiming1991}(circles) and Stassis et al.\cite{Stassis1978}(filled circles). The dotted vertical line is at $q=(\frac{2}{3},\frac{2}{3},\frac{2}{3})$ The observed softening at this point, experimental as well as theoretical, is important for the bcc--$\omega$ phase transition.}
\end{figure}

Looking at Zr in Fig. \ref{fig:zr_disprelations} we see once again excellent agreement of the results obtained by our method with experimental values. Worth noting is the near perfect location of the soft mode at the so-called $\omega$-point in the H-P direction. It is indicated by the red vertical line. This mode is of crucial importance for the bcc to $\omega$ phase transition in Zr and it has been difficult to reproduce in previous calculations\cite{Souvatzis2008a}.

\begin{figure}[th]
\includegraphics[width=8cm]{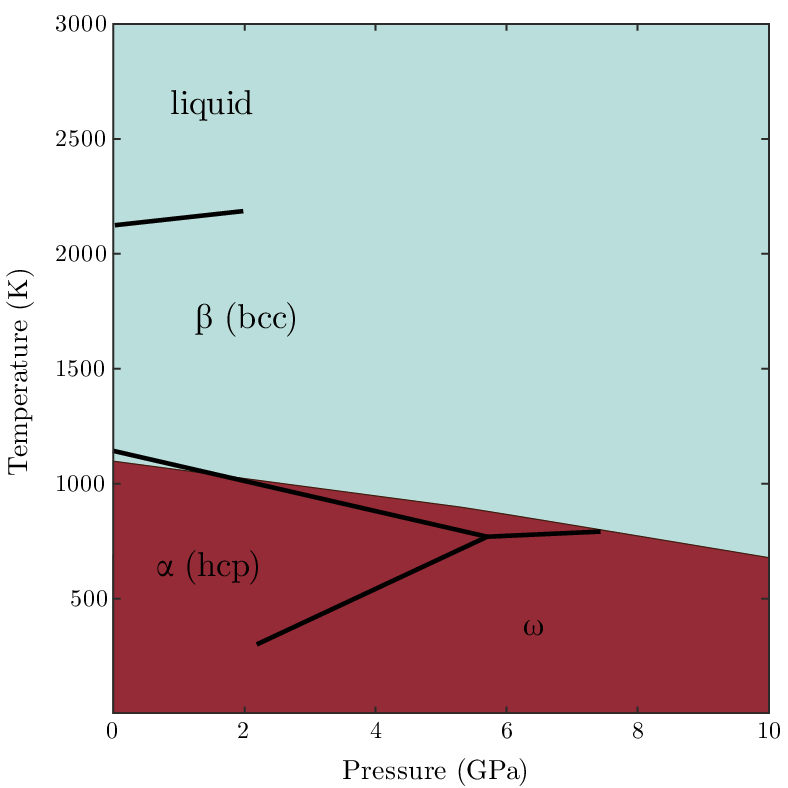}
\caption{\label{fig:phasediagram}(color online) Calculated bcc-hcp phase diagram for Zr. The calculated stability fields for low temperature hcp and high-temperature bcc phases are given in light blue (light grey) and red (dark grey) colors, respectively. The black lines show the experimental phase diagram\cite{Young1991a}, which also includes solid $\omega$ and liquid phases, not considered in simulations.}
\end{figure}

The phonon dispersion relations, while interesting, are not the main goal of this work. We want a solid method to deal with lattice dynamics for strongly anharmonic dynamically unstable systems. To test this we calculated the Gibbs free energy surface for bcc and hcp Zr, and in Fig. \ref{fig:phasediagram} we present the calculated bcc-hcp phase diagram. The dynamically unstable bcc free energies were calculated from the phonon dispersion relations as detailed above, on a grid of six volumes and five temperatures (volume $\pm 20\%$, temperature 100-1700K).

 For dynamically stable hcp phase we used the quasi-harmonic approximation\cite{Alfe2009} with phonon dispersion relations obtained on six volumes. We observe excellent agreement with experimentally determined line for the hcp to bcc transition over the whole interval of pressures and temperatures. 

In summary we have developed a thorough, accurate and easily extendable formalism to deal with lattice dynamics at high temperatures where the traditional quasiharmonic approximation fails. Excellent agreement with experimental results for bcc Li and bcc Zr indicates the usefulness of the method.

We would like to thank the Swedish Research Council, (VR) for financial support and the Swedish National Infrastructure for Computing (SNIC) for computational resources. Thanks to the Swedish Government Strategic Research Area Grant in Materials Science including Functional Materials. We are thankful to Prof. Yuri Vekilov for valuable discussions and Eyvas Isaev for invaluable expertise.


\begin{thebibliography}{32}%
\makeatletter
\providecommand \@ifxundefined [1]{%
 \@ifx{#1\undefined}
}%
\providecommand \@ifnum [1]{%
 \ifnum #1\expandafter \@firstoftwo
 \else \expandafter \@secondoftwo
 \fi
}%
\providecommand \@ifx [1]{%
 \ifx #1\expandafter \@firstoftwo
 \else \expandafter \@secondoftwo
 \fi
}%
\providecommand \natexlab [1]{#1}%
\providecommand \enquote  [1]{``#1''}%
\providecommand \bibnamefont  [1]{#1}%
\providecommand \bibfnamefont [1]{#1}%
\providecommand \citenamefont [1]{#1}%
\providecommand \href@noop [0]{\@secondoftwo}%
\providecommand \href [0]{\begingroup \@sanitize@url \@href}%
\providecommand \@href[1]{\@@startlink{#1}\@@href}%
\providecommand \@@href[1]{\endgroup#1\@@endlink}%
\providecommand \@sanitize@url [0]{\catcode `\\12\catcode `\$12\catcode
  `\&12\catcode `\#12\catcode `\^12\catcode `\_12\catcode `\%12\relax}%
\providecommand \@@startlink[1]{}%
\providecommand \@@endlink[0]{}%
\providecommand \url  [0]{\begingroup\@sanitize@url \@url }%
\providecommand \@url [1]{\endgroup\@href {#1}{\urlprefix }}%
\providecommand \urlprefix  [0]{URL }%
\providecommand \Eprint [0]{\href }%
\providecommand \doibase [0]{http://dx.doi.org/}%
\providecommand \selectlanguage [0]{\@gobble}%
\providecommand \bibinfo  [0]{\@secondoftwo}%
\providecommand \bibfield  [0]{\@secondoftwo}%
\providecommand \translation [1]{[#1]}%
\providecommand \BibitemOpen [0]{}%
\providecommand \bibitemStop [0]{}%
\providecommand \bibitemNoStop [0]{.\EOS\space}%
\providecommand \EOS [0]{\spacefactor3000\relax}%
\providecommand \BibitemShut  [1]{\csname bibitem#1\endcsname}%
\let\auto@bib@innerbib\@empty
\bibitem [{\citenamefont {Kohn}\ and\ \citenamefont {Sham}(1965)}]{Kohn1965c}%
  \BibitemOpen
  \bibfield  {author} {\bibinfo {author} {\bibfnamefont {W.}~\bibnamefont
  {Kohn}}\ and\ \bibinfo {author} {\bibfnamefont {L.~J.}\ \bibnamefont
  {Sham}},\ }\href {\doibase 10.1103/PhysRev.140.A1133} {\bibfield  {journal}
  {\bibinfo  {journal} {Physical Review}\ }\textbf {\bibinfo {volume} {140}},\
  \bibinfo {pages} {A1133} (\bibinfo {year} {1965})}\BibitemShut {NoStop}%
\bibitem [{\citenamefont {Fultz}(2010)}]{Fultz2010}%
  \BibitemOpen
  \bibfield  {author} {\bibinfo {author} {\bibfnamefont {B.}~\bibnamefont
  {Fultz}},\ }\href {\doibase 10.1016/j.pmatsci.2009.05.002} {\bibfield
  {journal} {\bibinfo  {journal} {Progress in Materials Science}\ }\textbf
  {\bibinfo {volume} {55}},\ \bibinfo {pages} {247} (\bibinfo {year}
  {2010})}\BibitemShut {NoStop}%
\bibitem [{\citenamefont {Baroni}\ \emph {et~al.}(2001)\citenamefont {Baroni},
  \citenamefont {de~Gironcoli},\ and\ \citenamefont {{Dal
  Corso}}}]{Baroni2001b}%
  \BibitemOpen
  \bibfield  {author} {\bibinfo {author} {\bibfnamefont {S.}~\bibnamefont
  {Baroni}}, \bibinfo {author} {\bibfnamefont {S.}~\bibnamefont
  {de~Gironcoli}}, \ and\ \bibinfo {author} {\bibfnamefont {A.}~\bibnamefont
  {{Dal Corso}}},\ }\href {\doibase 10.1103/RevModPhys.73.515} {\bibfield
  {journal} {\bibinfo  {journal} {Reviews of Modern Physics}\ }\textbf
  {\bibinfo {volume} {73}},\ \bibinfo {pages} {515} (\bibinfo {year}
  {2001})}\BibitemShut {NoStop}%
\bibitem [{\citenamefont {van~de Walle}\ and\ \citenamefont
  {Ceder}(2002)}]{VandeWalle2002}%
  \BibitemOpen
  \bibfield  {author} {\bibinfo {author} {\bibfnamefont {A.}~\bibnamefont
  {van~de Walle}}\ and\ \bibinfo {author} {\bibfnamefont {G.}~\bibnamefont
  {Ceder}},\ }\href {\doibase 10.1103/RevModPhys.74.11} {\bibfield  {journal}
  {\bibinfo  {journal} {Reviews of Modern Physics}\ }\textbf {\bibinfo {volume}
  {74}},\ \bibinfo {pages} {11} (\bibinfo {year} {2002})}\BibitemShut {NoStop}%
\bibitem [{\citenamefont {Dubrovinsky}\ \emph {et~al.}(2007)\citenamefont
  {Dubrovinsky}, \citenamefont {Dubrovinskaia}, \citenamefont {Narygina},
  \citenamefont {Kantor}, \citenamefont {Kuznetzov}, \citenamefont
  {Prakapenka}, \citenamefont {Vitos}, \citenamefont {Johansson}, \citenamefont
  {Mikhaylushkin}, \citenamefont {Simak},\ and\ \citenamefont
  {Abrikosov}}]{Dubrovinsky2007}%
  \BibitemOpen
  \bibfield  {author} {\bibinfo {author} {\bibfnamefont {L.}~\bibnamefont
  {Dubrovinsky}}, \bibinfo {author} {\bibfnamefont {N.}~\bibnamefont
  {Dubrovinskaia}}, \bibinfo {author} {\bibfnamefont {O.}~\bibnamefont
  {Narygina}}, \bibinfo {author} {\bibfnamefont {I.}~\bibnamefont {Kantor}},
  \bibinfo {author} {\bibfnamefont {A.}~\bibnamefont {Kuznetzov}}, \bibinfo
  {author} {\bibfnamefont {V.~B.}\ \bibnamefont {Prakapenka}}, \bibinfo
  {author} {\bibfnamefont {L.}~\bibnamefont {Vitos}}, \bibinfo {author}
  {\bibfnamefont {B.}~\bibnamefont {Johansson}}, \bibinfo {author}
  {\bibfnamefont {A.~S.}\ \bibnamefont {Mikhaylushkin}}, \bibinfo {author}
  {\bibfnamefont {S.~I.}\ \bibnamefont {Simak}}, \ and\ \bibinfo {author}
  {\bibfnamefont {I.~A.}\ \bibnamefont {Abrikosov}},\ }\href {\doibase
  10.1126/science.1142105} {\bibfield  {journal} {\bibinfo  {journal} {Science
  (New York, N.Y.)}\ }\textbf {\bibinfo {volume} {316}},\ \bibinfo {pages}
  {1880} (\bibinfo {year} {2007})}\BibitemShut {NoStop}%
\bibitem [{\citenamefont {Klein}\ and\ \citenamefont
  {Horton}(1972)}]{Klein1972}%
  \BibitemOpen
  \bibfield  {author} {\bibinfo {author} {\bibfnamefont {M.~L.}\ \bibnamefont
  {Klein}}\ and\ \bibinfo {author} {\bibfnamefont {G.~K.}\ \bibnamefont
  {Horton}},\ }\href {\doibase 10.1007/BF00654839} {\bibfield  {journal}
  {\bibinfo  {journal} {Journal of Low Temperature Physics}\ }\textbf {\bibinfo
  {volume} {9}},\ \bibinfo {pages} {151} (\bibinfo {year} {1972})}\BibitemShut
  {NoStop}%
\bibitem [{\citenamefont {Born}(1951)}]{Born1951}%
  \BibitemOpen
  \bibfield  {author} {\bibinfo {author} {\bibfnamefont {M.}~\bibnamefont
  {Born}},\ }\href@noop {} {\emph {\bibinfo {title} {{Fest. Akad. D. Wiss,
  G\"{o}ttingen, Math. Phys. Klasse}}}}\ (\bibinfo  {publisher} {Springer},\
  \bibinfo {address} {Berlin},\ \bibinfo {year} {1951})\BibitemShut {NoStop}%
\bibitem [{\citenamefont {Hooton}(1955)}]{Hooton}%
  \BibitemOpen
  \bibfield  {author} {\bibinfo {author} {\bibfnamefont {D.~J.}\ \bibnamefont
  {Hooton}},\ }\href {\doibase 10.1007/BF01330055} {\bibfield  {journal}
  {\bibinfo  {journal} {Zeitschrift f\"{u}r Physik}\ }\textbf {\bibinfo
  {volume} {142}},\ \bibinfo {pages} {42} (\bibinfo {year} {1955})}\BibitemShut
  {NoStop}%
\bibitem [{\citenamefont {Born}\ and\ \citenamefont {Hooton}(1955)}]{Born}%
  \BibitemOpen
  \bibfield  {author} {\bibinfo {author} {\bibfnamefont {M.}~\bibnamefont
  {Born}}\ and\ \bibinfo {author} {\bibfnamefont {D.~J.}\ \bibnamefont
  {Hooton}},\ }\href {\doibase 10.1007/BF01329422} {\bibfield  {journal}
  {\bibinfo  {journal} {Zeitschrift f\"{u}r Physik}\ }\textbf {\bibinfo
  {volume} {142}},\ \bibinfo {pages} {201} (\bibinfo {year}
  {1955})}\BibitemShut {NoStop}%
\bibitem [{\citenamefont {Choquard}(1967)}]{choquard1967}%
  \BibitemOpen
  \bibfield  {author} {\bibinfo {author} {\bibfnamefont {P.}~\bibnamefont
  {Choquard}},\ }\href@noop {} {\emph {\bibinfo {title} {{The anharmonic
  crystal}}}}\ (\bibinfo  {publisher} {W. A. Benjamin, INC.},\ \bibinfo
  {address} {New York},\ \bibinfo {year} {1967})\BibitemShut {NoStop}%
\bibitem [{\citenamefont {Plakida}(1970)}]{Plakida1970}%
  \BibitemOpen
  \bibfield  {author} {\bibinfo {author} {\bibfnamefont {N.~M.}\ \bibnamefont
  {Plakida}},\ }\href {\doibase 10.1007/BF01035987} {\bibfield  {journal}
  {\bibinfo  {journal} {Theoretical and Mathematical Physics}\ }\textbf
  {\bibinfo {volume} {5}},\ \bibinfo {pages} {1047} (\bibinfo {year}
  {1970})}\BibitemShut {NoStop}%
\bibitem [{\citenamefont {Souvatzis}\ \emph {et~al.}(2008)\citenamefont
  {Souvatzis}, \citenamefont {Eriksson}, \citenamefont {Katsnelson},\ and\
  \citenamefont {Rudin}}]{Souvatzis2008a}%
  \BibitemOpen
  \bibfield  {author} {\bibinfo {author} {\bibfnamefont {P.}~\bibnamefont
  {Souvatzis}}, \bibinfo {author} {\bibfnamefont {O.}~\bibnamefont {Eriksson}},
  \bibinfo {author} {\bibfnamefont {M.}~\bibnamefont {Katsnelson}}, \ and\
  \bibinfo {author} {\bibfnamefont {S.}~\bibnamefont {Rudin}},\ }\href
  {\doibase 10.1103/PhysRevLett.100.095901} {\bibfield  {journal} {\bibinfo
  {journal} {Physical Review Letters}\ }\textbf {\bibinfo {volume} {100}},\
  \bibinfo {pages} {95901} (\bibinfo {year} {2008})}\BibitemShut {NoStop}%
\bibitem [{\citenamefont {Souvatzis}\ \emph {et~al.}(2009)\citenamefont
  {Souvatzis}, \citenamefont {Eriksson}, \citenamefont {Katsnelson},\ and\
  \citenamefont {Rudin}}]{Souvatzis2009}%
  \BibitemOpen
  \bibfield  {author} {\bibinfo {author} {\bibfnamefont {P.}~\bibnamefont
  {Souvatzis}}, \bibinfo {author} {\bibfnamefont {O.}~\bibnamefont {Eriksson}},
  \bibinfo {author} {\bibfnamefont {M.}~\bibnamefont {Katsnelson}}, \ and\
  \bibinfo {author} {\bibfnamefont {S.}~\bibnamefont {Rudin}},\ }\href
  {\doibase 10.1016/j.commatsci.2008.06.016} {\bibfield  {journal} {\bibinfo
  {journal} {Computational Materials Science}\ }\textbf {\bibinfo {volume}
  {44}},\ \bibinfo {pages} {888} (\bibinfo {year} {2009})}\BibitemShut
  {NoStop}%
\bibitem [{\citenamefont {Dickey}\ and\ \citenamefont
  {Paskin}(1969)}]{Dickey1969}%
  \BibitemOpen
  \bibfield  {author} {\bibinfo {author} {\bibfnamefont {J.}~\bibnamefont
  {Dickey}}\ and\ \bibinfo {author} {\bibfnamefont {A.}~\bibnamefont
  {Paskin}},\ }\href {\doibase 10.1103/PhysRev.188.1407} {\bibfield  {journal}
  {\bibinfo  {journal} {Physical Review}\ }\textbf {\bibinfo {volume} {188}},\
  \bibinfo {pages} {1407} (\bibinfo {year} {1969})}\BibitemShut {NoStop}%
\bibitem [{\citenamefont {Vo\v{c}adlo}\ and\ \citenamefont
  {Alf\`{e}}(2002)}]{Vocadlo2002}%
  \BibitemOpen
  \bibfield  {author} {\bibinfo {author} {\bibfnamefont {L.}~\bibnamefont
  {Vo\v{c}adlo}}\ and\ \bibinfo {author} {\bibfnamefont {D.}~\bibnamefont
  {Alf\`{e}}},\ }\href {\doibase 10.1103/PhysRevB.65.214105} {\bibfield
  {journal} {\bibinfo  {journal} {Physical Review B}\ }\textbf {\bibinfo
  {volume} {65}},\ \bibinfo {pages} {214105} (\bibinfo {year}
  {2002})}\BibitemShut {NoStop}%
\bibitem [{\citenamefont {Grabowski}\ \emph {et~al.}(2009)\citenamefont
  {Grabowski}, \citenamefont {Ismer}, \citenamefont {Hickel},\ and\
  \citenamefont {Neugebauer}}]{Grabowski2009c}%
  \BibitemOpen
  \bibfield  {author} {\bibinfo {author} {\bibfnamefont {B.}~\bibnamefont
  {Grabowski}}, \bibinfo {author} {\bibfnamefont {L.}~\bibnamefont {Ismer}},
  \bibinfo {author} {\bibfnamefont {T.}~\bibnamefont {Hickel}}, \ and\ \bibinfo
  {author} {\bibfnamefont {J.}~\bibnamefont {Neugebauer}},\ }\href {\doibase
  10.1103/PhysRevB.79.134106} {\bibfield  {journal} {\bibinfo  {journal}
  {Physical Review B}\ }\textbf {\bibinfo {volume} {79}},\ \bibinfo {pages}
  {134106} (\bibinfo {year} {2009})}\BibitemShut {NoStop}%
\bibitem [{\citenamefont {Alf\`{e}}(2009)}]{Alfe2009}%
  \BibitemOpen
  \bibfield  {author} {\bibinfo {author} {\bibfnamefont {D.}~\bibnamefont
  {Alf\`{e}}},\ }\href {\doibase 10.1016/j.cpc.2009.03.010} {\bibfield
  {journal} {\bibinfo  {journal} {Computer Physics Communications}\ }\textbf
  {\bibinfo {volume} {180}},\ \bibinfo {pages} {2622} (\bibinfo {year}
  {2009})}\BibitemShut {NoStop}%
\bibitem [{Note1()}]{Note1}%
  \BibitemOpen
  \bibinfo {note} {In this case $F_L$ are from a 128 atom bcc Li supercell,
  $2\times 2 \times 2$ kpoint grid, and $F^H$ from the same configuration, with
  50\% higher energy cutoff and $7\times 7 \times 7$ kpoint grid.}\BibitemShut
  {Stop}%
\bibitem [{\citenamefont {Overhauser}(1984)}]{Overhauser1984}%
  \BibitemOpen
  \bibfield  {author} {\bibinfo {author} {\bibfnamefont {A.}~\bibnamefont
  {Overhauser}},\ }\href {\doibase 10.1103/PhysRevLett.53.64} {\bibfield
  {journal} {\bibinfo  {journal} {Physical Review Letters}\ }\textbf {\bibinfo
  {volume} {53}},\ \bibinfo {pages} {64} (\bibinfo {year} {1984})}\BibitemShut
  {NoStop}%
\bibitem [{\citenamefont {Neaton}\ and\ \citenamefont
  {Ashcroft}(1999)}]{Neaton1999a}%
  \BibitemOpen
  \bibfield  {author} {\bibinfo {author} {\bibfnamefont {J.~B.}\ \bibnamefont
  {Neaton}}\ and\ \bibinfo {author} {\bibfnamefont {N.~W.}\ \bibnamefont
  {Ashcroft}},\ }\href {\doibase 10.1038/22067} {\bibfield  {journal} {\bibinfo
   {journal} {Nature}\ }\textbf {\bibinfo {volume} {400}},\ \bibinfo {pages}
  {141} (\bibinfo {year} {1999})}\BibitemShut {NoStop}%
\bibitem [{\citenamefont {Ye}\ \emph {et~al.}(1987)\citenamefont {Ye},
  \citenamefont {Chen}, \citenamefont {Ho}, \citenamefont {Harmon},\ and\
  \citenamefont {Lindgrd}}]{Ye1987}%
  \BibitemOpen
  \bibfield  {author} {\bibinfo {author} {\bibfnamefont {Y.}~\bibnamefont
  {Ye}}, \bibinfo {author} {\bibfnamefont {Y.}~\bibnamefont {Chen}}, \bibinfo
  {author} {\bibfnamefont {K.}~\bibnamefont {Ho}}, \bibinfo {author}
  {\bibfnamefont {B.}~\bibnamefont {Harmon}}, \ and\ \bibinfo {author}
  {\bibfnamefont {P.}~\bibnamefont {Lindgrd}},\ }\href {\doibase
  10.1103/PhysRevLett.58.1769} {\bibfield  {journal} {\bibinfo  {journal}
  {Physical Review Letters}\ }\textbf {\bibinfo {volume} {58}},\ \bibinfo
  {pages} {1769} (\bibinfo {year} {1987})}\BibitemShut {NoStop}%
\bibitem [{\citenamefont {Kresse}(1999)}]{Kresse1999}%
  \BibitemOpen
  \bibfield  {author} {\bibinfo {author} {\bibfnamefont {G.}~\bibnamefont
  {Kresse}},\ }\href {\doibase 10.1103/PhysRevB.59.1758} {\bibfield  {journal}
  {\bibinfo  {journal} {Physical Review B}\ }\textbf {\bibinfo {volume} {59}},\
  \bibinfo {pages} {1758} (\bibinfo {year} {1999})}\BibitemShut {NoStop}%
\bibitem [{\citenamefont {Kresse}\ and\ \citenamefont
  {Furthm\"{u}ller}(1996)}]{Kresse1996}%
  \BibitemOpen
  \bibfield  {author} {\bibinfo {author} {\bibfnamefont {G.}~\bibnamefont
  {Kresse}}\ and\ \bibinfo {author} {\bibfnamefont {J.}~\bibnamefont
  {Furthm\"{u}ller}},\ }\href {\doibase 10.1103/PhysRevB.54.11169} {\bibfield
  {journal} {\bibinfo  {journal} {Physical Review B}\ }\textbf {\bibinfo
  {volume} {54}},\ \bibinfo {pages} {11169} (\bibinfo {year}
  {1996})}\BibitemShut {NoStop}%
\bibitem [{\citenamefont {Kresse}\ and\ \citenamefont
  {Hafner}(1993)}]{Kresse1993b}%
  \BibitemOpen
  \bibfield  {author} {\bibinfo {author} {\bibfnamefont {G.}~\bibnamefont
  {Kresse}}\ and\ \bibinfo {author} {\bibfnamefont {J.}~\bibnamefont
  {Hafner}},\ }\href {\doibase 10.1103/PhysRevB.48.13115} {\bibfield  {journal}
  {\bibinfo  {journal} {Physical Review B}\ }\textbf {\bibinfo {volume} {48}},\
  \bibinfo {pages} {13115} (\bibinfo {year} {1993})}\BibitemShut {NoStop}%
\bibitem [{\citenamefont {Kresse}(1996)}]{Kresse1996c}%
  \BibitemOpen
  \bibfield  {author} {\bibinfo {author} {\bibfnamefont {G.}~\bibnamefont
  {Kresse}},\ }\href {\doibase 10.1016/0927-0256(96)00008-0} {\bibfield
  {journal} {\bibinfo  {journal} {Computational Materials Science}\ }\textbf
  {\bibinfo {volume} {6}},\ \bibinfo {pages} {15} (\bibinfo {year}
  {1996})}\BibitemShut {NoStop}%
\bibitem [{\citenamefont {Perdew}\ \emph {et~al.}(1996)\citenamefont {Perdew},
  \citenamefont {Burke},\ and\ \citenamefont {Ernzerhof}}]{Perdew1996}%
  \BibitemOpen
  \bibfield  {author} {\bibinfo {author} {\bibfnamefont {J.~P.}\ \bibnamefont
  {Perdew}}, \bibinfo {author} {\bibfnamefont {K.}~\bibnamefont {Burke}}, \
  and\ \bibinfo {author} {\bibfnamefont {M.}~\bibnamefont {Ernzerhof}},\ }\href
  {\doibase 10.1103/PhysRevLett.77.3865} {\bibfield  {journal} {\bibinfo
  {journal} {Physical Review Letters}\ }\textbf {\bibinfo {volume} {77}},\
  \bibinfo {pages} {3865} (\bibinfo {year} {1996})}\BibitemShut {NoStop}%
\bibitem [{\citenamefont {Hoover}(1985)}]{Hoover1985}%
  \BibitemOpen
  \bibfield  {author} {\bibinfo {author} {\bibfnamefont {W.}~\bibnamefont
  {Hoover}},\ }\href {\doibase 10.1103/PhysRevA.31.1695} {\bibfield  {journal}
  {\bibinfo  {journal} {Physical Review A}\ }\textbf {\bibinfo {volume} {31}},\
  \bibinfo {pages} {1695} (\bibinfo {year} {1985})}\BibitemShut {NoStop}%
\bibitem [{\citenamefont {Nos\'{e}}(1984)}]{Nose1984}%
  \BibitemOpen
  \bibfield  {author} {\bibinfo {author} {\bibfnamefont {S.}~\bibnamefont
  {Nos\'{e}}},\ }\href {\doibase 10.1080/00268978400101201} {\bibfield
  {journal} {\bibinfo  {journal} {Molecular Physics}\ }\textbf {\bibinfo
  {volume} {52}},\ \bibinfo {pages} {255} (\bibinfo {year} {1984})}\BibitemShut
  {NoStop}%
\bibitem [{\citenamefont {Beg}\ and\ \citenamefont {Nielsen}(1976)}]{Beg1976}%
  \BibitemOpen
  \bibfield  {author} {\bibinfo {author} {\bibfnamefont {M.}~\bibnamefont
  {Beg}}\ and\ \bibinfo {author} {\bibfnamefont {M.}~\bibnamefont {Nielsen}},\
  }\href {\doibase 10.1103/PhysRevB.14.4266} {\bibfield  {journal} {\bibinfo
  {journal} {Physical Review B}\ }\textbf {\bibinfo {volume} {14}},\ \bibinfo
  {pages} {4266} (\bibinfo {year} {1976})}\BibitemShut {NoStop}%
\bibitem [{\citenamefont {Heiming}\ \emph {et~al.}(1991)\citenamefont
  {Heiming}, \citenamefont {Petry}, \citenamefont {Trampenau}, \citenamefont
  {Alba}, \citenamefont {Herzig}, \citenamefont {Schober},\ and\ \citenamefont
  {Vogl}}]{Heiming1991}%
  \BibitemOpen
  \bibfield  {author} {\bibinfo {author} {\bibfnamefont {A.}~\bibnamefont
  {Heiming}}, \bibinfo {author} {\bibfnamefont {W.}~\bibnamefont {Petry}},
  \bibinfo {author} {\bibfnamefont {J.}~\bibnamefont {Trampenau}}, \bibinfo
  {author} {\bibfnamefont {M.}~\bibnamefont {Alba}}, \bibinfo {author}
  {\bibfnamefont {C.}~\bibnamefont {Herzig}}, \bibinfo {author} {\bibfnamefont
  {H.}~\bibnamefont {Schober}}, \ and\ \bibinfo {author} {\bibfnamefont
  {G.}~\bibnamefont {Vogl}},\ }\href {\doibase 10.1103/PhysRevB.43.10948}
  {\bibfield  {journal} {\bibinfo  {journal} {Physical Review B}\ }\textbf
  {\bibinfo {volume} {43}},\ \bibinfo {pages} {10948} (\bibinfo {year}
  {1991})}\BibitemShut {NoStop}%
\bibitem [{\citenamefont {Stassis}\ \emph {et~al.}(1978)\citenamefont
  {Stassis}, \citenamefont {Zarestky},\ and\ \citenamefont
  {Wakabayashi}}]{Stassis1978}%
  \BibitemOpen
  \bibfield  {author} {\bibinfo {author} {\bibfnamefont {C.}~\bibnamefont
  {Stassis}}, \bibinfo {author} {\bibfnamefont {J.}~\bibnamefont {Zarestky}}, \
  and\ \bibinfo {author} {\bibfnamefont {N.}~\bibnamefont {Wakabayashi}},\
  }\href {\doibase 10.1103/PhysRevLett.41.1726} {\bibfield  {journal} {\bibinfo
   {journal} {Physical Review Letters}\ }\textbf {\bibinfo {volume} {41}},\
  \bibinfo {pages} {1726} (\bibinfo {year} {1978})}\BibitemShut {NoStop}%
\bibitem [{\citenamefont {Young}(1991)}]{Young1991a}%
  \BibitemOpen
  \bibfield  {author} {\bibinfo {author} {\bibfnamefont {D.}~\bibnamefont
  {Young}},\ }\href@noop {} {\emph {\bibinfo {title} {{Phase diagrams of the
  elements}}}}\ (\bibinfo  {publisher} {University of California Press},\
  \bibinfo {year} {1991})\BibitemShut {NoStop}%
\end{thebibliography}

%

\end{document}